\documentclass{iau379}

\usepackage{graphicx}
\usepackage{amsmath}
\usepackage{multirow}

\newcommand{\hi}{H\textsc{i}}
\newcommand{\spexxy}{\texttt{spexxy}}

\begin{document}

\lefttitle{Vaz and Brinchmann}
\righttitle{IAU Symposium 379: Leo T Dissected with the MUSE-Faint Survey}

\jnlPage{1}{7}
\jnlDoiYr{2023}
\doival{10.1017/xxxxx}

\aopheadtitle{Proceedings of IAU Symposium 379}
\editors{P. Bonifacio,  M.-R. Cioni, F. Hammer, M. Pawlowski, and S. Taibi, eds.}

\title{Leo T Dissected with the MUSE-Faint Survey}

\author{Daniel Vaz $^{1,2}$, Jarle Brinchmann$^{1,3}$ \& The MUSE Collaboration}
\affiliation{$^1$ Instituto de Astrofísica e Ciências do Espaço, Universidade do Porto, CAUP, Rua das Estrelas, 4150-762 Porto, Portugal\\
$^2$ Departamento de Física e Astronomia, Faculdade de Ciências, Universidade do Porto, Rua do Campo Alegre 687, PT4169-007 Porto, Portugal\\
$^3$ Leiden Observatory, Leiden University, PO Box 9513, 2300 RA Leiden, The Netherlands }

\begin{abstract}

Leo T is the lowest mass galaxy known to contain neutral gas and to show signs of recent star formation, which makes it a valuable laboratory for studying the nature of gas and star formation at the limits of where galaxies are found to have rejuvenating episodes of star formation.

Here we discuss a novel study of Leo T that uses data from the MUSE integral field spectrograph and photometric data from HST. The high sensitivity of MUSE allowed us to increase the number of Leo T stars observed spectroscopically from 19 to 75. We studied the age and metallicity of these stars and identified two populations, all consistent with similar metallicity of [Fe/H] $\sim$ -1.5 dex, suggesting that a large fraction of metals were ejected. Within the young population, we discovered three emission line Be stars, supporting the conclusion that rapidly rotating massive stars are common in metal-poor environments.
We find differences in the dynamics of young and old stars, with the young population having a velocity dispersion consistent with the kinematics of the cold component of the neutral gas. This finding directly links the recent star formation in Leo T with the cold component of the neutral gas.

\end{abstract}


\maketitle
\vspace{-15pt}
\section{Introduction}

Ultra-Faint Dwarf galaxies (UFDs) represent a fascinating enigma in the study of the Universe. These elusive objects are characterised by their extremely low metallicities, simple assembly histories, and dominant dark matter content \citep{Simon2019}, making them an essential piece of the puzzle in understanding galaxy formation and evolution.

Among the faint and ultra-faint dwarf sample, Leo T stands out as a particularly intriguing object that has received significant attention from astronomers. Leo T is the faintest and least massive dwarf galaxy known to contain neutral gas and exhibit signs of recent star formation. This unique set of characteristics makes Leo T a valuable testing ground for galaxy formation models, as they present a challenge to current theories that have struggled to reproduce similar galaxies. Further observations of Leo T will enable astronomers to refine their models and determine whether they are on the right track towards a comprehensive and predictive theory of galaxy formation.

Leo T was discovered using SDSS imaging by \cite{Irwin2007}. The stellar mass of Leo T is estimated to be $M~=~1.3~\times 10 ^ 5 ~\mathrm{M_\odot}$ \citep{McConnachie2012}. The star formation history (SFH) of Leo T has been extensively studied \citep{Irwin2007, deJong2008, Weisz2012, Clementini2012}. 
These studies show that 50\% of the total stellar mass was formed prior to $7.6$ Gyr ago, with star formation beginning over 10 Gyr ago and continuing until recent times. 
They also show evidence of a quenching of star formation in Leo T about $25$ Myr ago. None of the studies found evidence of an evolution in isochronal metallicity such that, over the course of its lifetime, it is consistent with a constant value of $[M/H]\sim~-1.6$.

The only previous spectroscopic observations of Leo T are
those of \cite{Simon2007}. They derive a mean radial velocity of $v_{rad} = 38.1 ~\pm 2 ~\mathrm{km~ s^{-1}}$, and velocity dispersion of $\sigma_{v_{rad}} = 7.5 ~\pm 1.6 ~\mathrm{km~ s^{-1}}$, which corresponds to a total dynamical mass of $8.2~\times 10 ^ 6 ~\mathrm{M_\odot}$.

\cite{Weber2008} and \cite{AdamsOosterloo2018} concluded that Leo T contains neutral gas. The \hi\ mass is almost four times the stellar mass, with $M_{\hi}~=~4.1~\times 10 ^ 5 ~\mathrm{M_\odot}$ \citep{AdamsOosterloo2018}. They show that the gas is present in a Cold Neutral Medium (CNM) (T $\sim$ 800K) and a Warm Neutral Medium (WNM) (T $\sim$ 6000K) \hi, with the CNM corresponding to almost 10\% of the total \hi\ mass.
Relevantly, the CNM was found to have a lower velocity dispersion ($\sigma_{\mathrm{CNM}} = 2.5~\pm 0.1~\mathrm{km\ s^{-1}}$) than the WNM ($\sigma_{\mathrm{WNM}} = 7.1~\pm 0.4~\mathrm{km\ s^{-1}}$) \citep{AdamsOosterloo2018}. 
The presence of this large cold component raises the question of whether this component is related to the recent star formation observed in Leo T.

To further investigate Leo T, we used spectroscopic observations using the Multi-Unit Spectroscopic Explorer (MUSE, \citealt{Bacon2010}) integral field spectrograph (IFS).
Succinctly, we densely map the stellar content and use the stellar spectra to measure the stellar metallicity and stellar kinematics. We find identifiers of a young population, namely Be stars.
The data and results presented here are discussed further in \cite{Zoutendijk2021b} and \cite{Vaz2023}. 

\section{Results and Discussion}

\begin{figure}[]
    \centering
  \includegraphics[scale=.32]{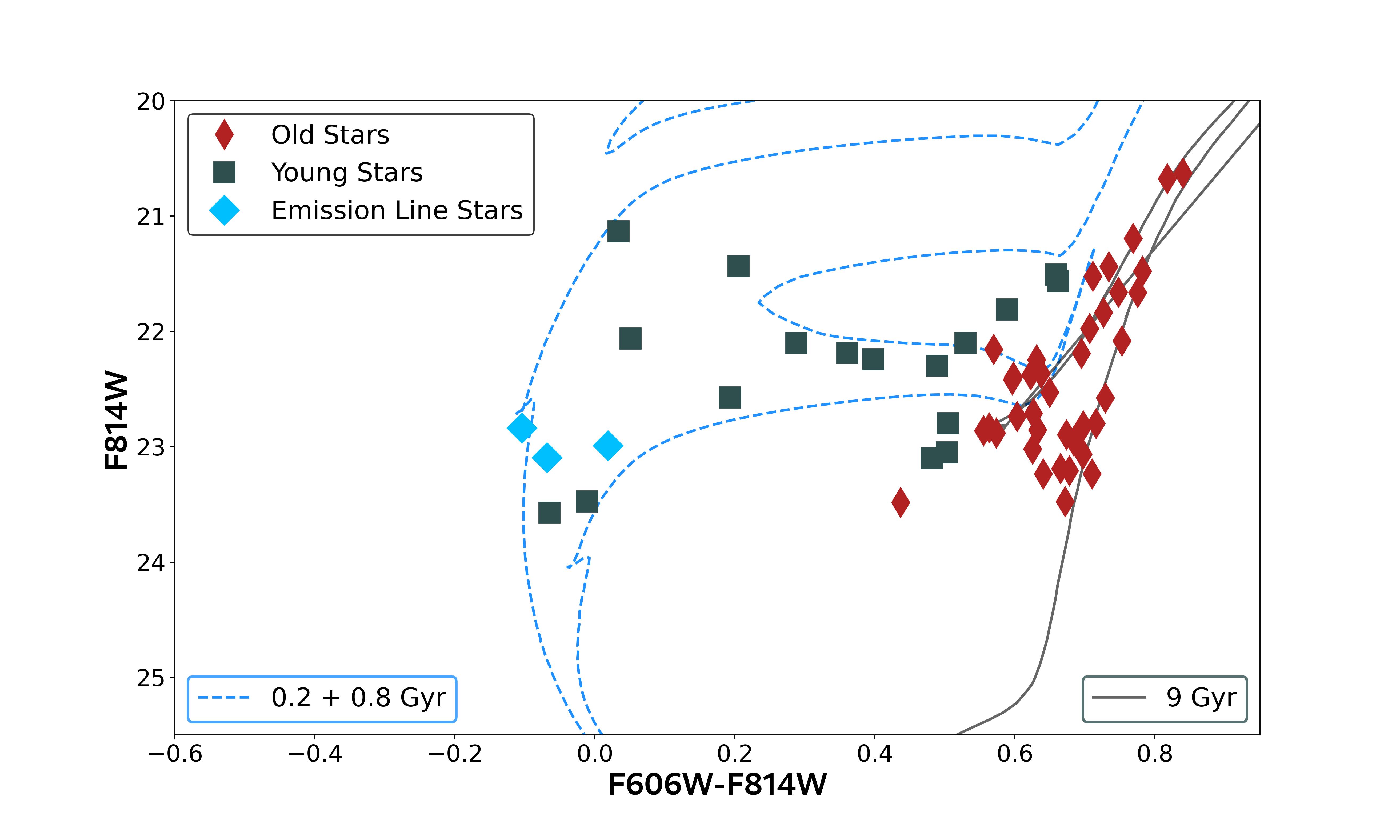}
  \caption{Color-magnitude diagram of the 58 Leo T stars detected with MUSE, plotted against PARSEC isochrones drawn for constant $[\mathrm{Fe/H}] = -1.6$. Three representative isochrones are plotted, two in blue, with ages of 0.2 and 0.8 Gyr, and one in gray for age of 9 Gyr. The stars that were found consistent with the younger isochrones are shown as dark blue squares, with the emission line stars shown as blue squares. The stars consistent with the older isochrones are shown as red diamonds.}
  \label{color_plot}
\end{figure}

\begin{figure}[]
    \centering
  \includegraphics[scale=.09]{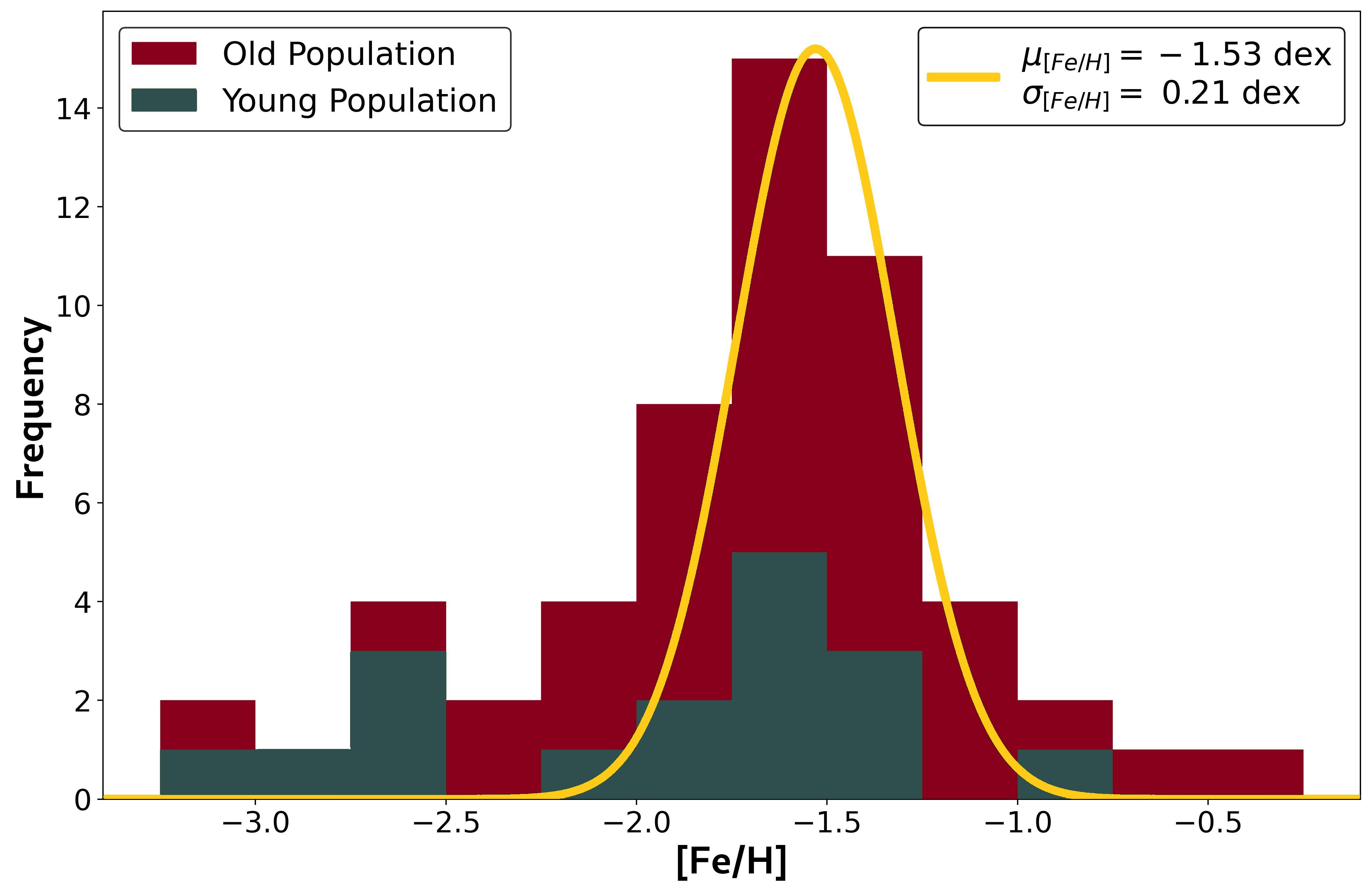}
  \caption{The distribution of the metallicity ([Fe/H]) of 55 Leo T stars, estimated using \spexxy. The younger population, consisting of 17 stars, is represented with a lighter color. Also plotted is the result of a MCMC fit for the mean and standard deviation of the distribution.}
  \label{his_met}
\end{figure}

The central region of Leo T was observed as part of MUSE-Faint \citep{Zoutendijk2020}, a MUSE GTO survey of UFDs (PI Brinchmann). MUSE \citep{Bacon2010} is a large-field medium-spectral-resolution integrated field spectrograph installed at Unit Telescope 4 of the Very Large Telescope (VLT). 
We extracted spectra from the final data cube using PampleMuse \citep{Kamann2013}. As a general rule, the extracted spectra have a modest signal-to-noise ratio (S/N) and spectral resolution (R $\sim 3000$). We used the \texttt{spexxy} full-spectrum fitting code \citep{Husser2016} together with the PHOENIX \citep{Husser2013} model spectra to estimate the physical parameters, namely line-of-sight velocities and [Fe/H]. 

\subsection{Different Populations in Leo T}

We were able to identify 58 stars as members of Leo T based on their kinematics. For 55 of these stars we were also able to obtain an estimate of [Fe/H]. The three stars without [Fe/H] estimates are emission line Be stars, which are discussed further below. By combining with the data from \cite{Simon2007} we have measurements of the kinematics for 75 stars.

We complemented this data with HST ACS F606W and F814W photometry\footnote{HST Proposals 12914, Principal Investigator Tuan Do and 14224, Principal Investigator Carmen Gallart}. 
We fit PARSEC stellar isochrones \citep{Bressan2012} to the resulting colour magnitude diagram for the 58 stars (Figure~\ref{color_plot}). 
The best-fit [M/H] for the isochrone is [M/H]=-1.6, which is consistent with the value found in \cite{Weisz2012}, and therefore we use this value to draw the representative isochrones shown in Figure~\ref{color_plot}. The first clear conclusion is that the sample covers a wide range of ages, with some stars consistent with ages as high as $> 10$ Gyr and others as low as $200$ Myr. As such, we divided the stars into two populations: a young population, of 20 stars consistent with ages $< 1$ Gyr, and an old population, of 38 stars consistent with ages $> 5$ Gyr. Both populations are displayed using different colors in Figure~\ref{color_plot}. To assign each star to a population, we covered the color magnitude space with isochrones of different ages and with [M/H]=-1.6. We assign each star the age of the nearest isochrone. Because there is a degeneracy between the different isochrones in certain parts of the color-color space, we repeated all analyses discussed below by reassign the stars that fall in a degenerated region. We find that this does not affect our results.

Within the young population we identified three emission line stars. We tentatively identified these as Be stars. Due to their peculiar spectra, \texttt{spexxy} failed to fit for metallicity and, therefore, these stars were not included in our metallicity analysis. Of relevance is the fact that they make up $15\%$ of the young sample, which is comparable to Milky Way studies that reported rates at a level of $10-20\%$ in stellar clusters \citep{Mathew2008}. More recent studies, such as \cite{Schootemeijer2022}, show that OBe stars and, therefore, rapidly rotating massive stars, are common in metal-poor environments, and the detection of Be stars in Leo T supports this conclusion and extend them to even lower metallicity.

\subsection{Chemical Evolution of Leo T}

The histogram of the metallicity estimates for the 55 stars is shown in Figure~\ref{his_met}. To quantify this distribution, we implemented an MCMC model, assuming that the underlying distribution is a Gaussian. Therefore, we fit the mean metallicity and the metallicity dispersion of the distribution, which are also shown in Figure~\ref{his_met}. We obtained a metallicity of $\mathrm{[Fe/H]} = -1.53 \pm 0.05$, which is in good agreement with our photometric analysis. We repeated the analysis by resampling and removing outliers, without consequences on the results (the outliers are usually low S/N).
We find a metallicity dispersion of $\sigma_{\mathrm{[Fe/H]}}= 0.21 \pm 0.06$, which is low, implying that all stars have similar metallicity and that Leo T underwent almost no metallicity evolution throughout its history. This, in conjunction with the extended history of star formation of Leo T, suggests that a large fraction of metals have been ejected, keeping metallicity constant. In fact, this is consistent with theoretical expectations for low-mass dwarf galaxies \citep{Emerick2018}.

We repeated this analysis separately for each population. Even though the results are consistent with each other, the results are not conclusive because the sample of young stars is too small (consisting only of 17 stars). In addition, the distribution is asymmetric, especially for the young stars, with the low S/N spectra preferring a lower metallicity, meaning that our constraints prefer a somewhat lower metallicity for the younger population. However, the uncertainties here are too high to draw any conclusions. 

\begin{figure}[]
    \centering
  \includegraphics[scale=.085]{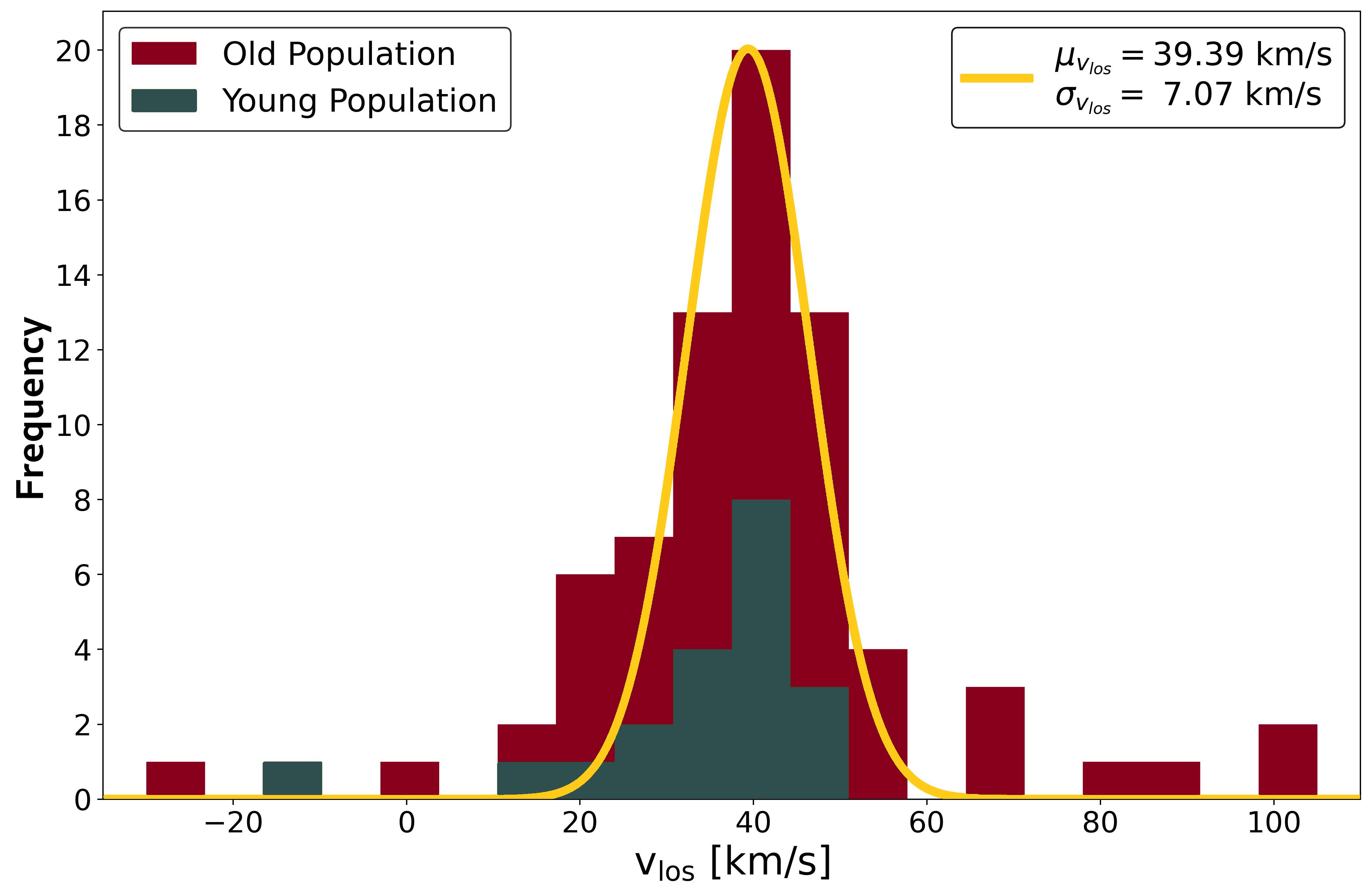}
  \caption{The distribution of the line-of-sight velocity of 75 Leo T stars, estimated using \spexxy. The younger population, consisting of 20 stars, is represented with a lighter color. Also plotted is the result of a MCMC fit for the mean and standard deviation of the distribution.}
  \label{hist_vel}
\end{figure}

\begin{figure}[]
    \centering
  \includegraphics[scale=.085]{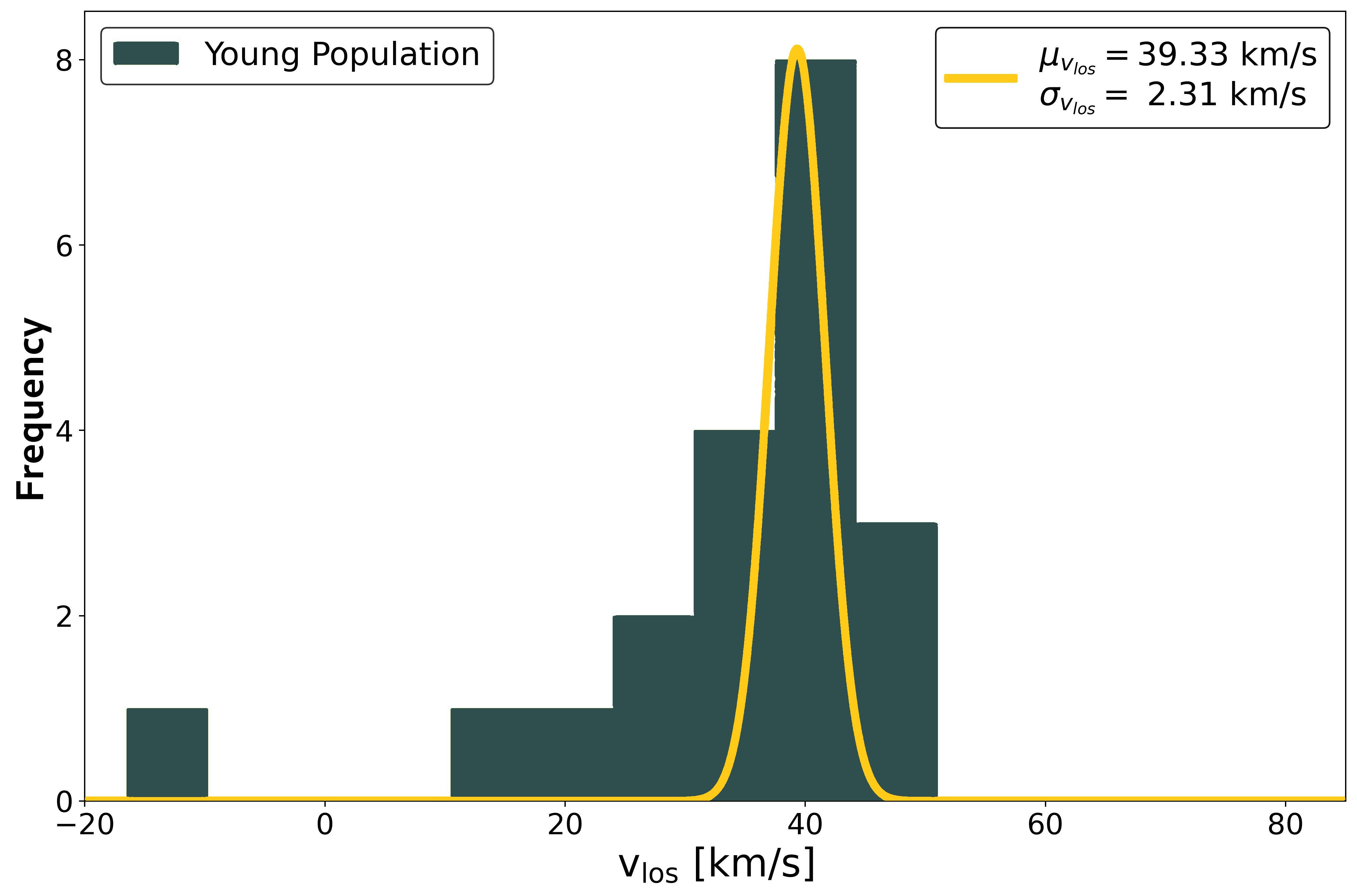}
  \caption{The distribution of the line-of-sight velocity of 20 young Leo T stars, estimated using \spexxy. Also plotted is the result of a MCMC fit for the mean and standard deviation of the distribution.}
  \label{hist_young}
\end{figure}

\begin{figure}[]
    \centering
  \includegraphics[scale=.09]{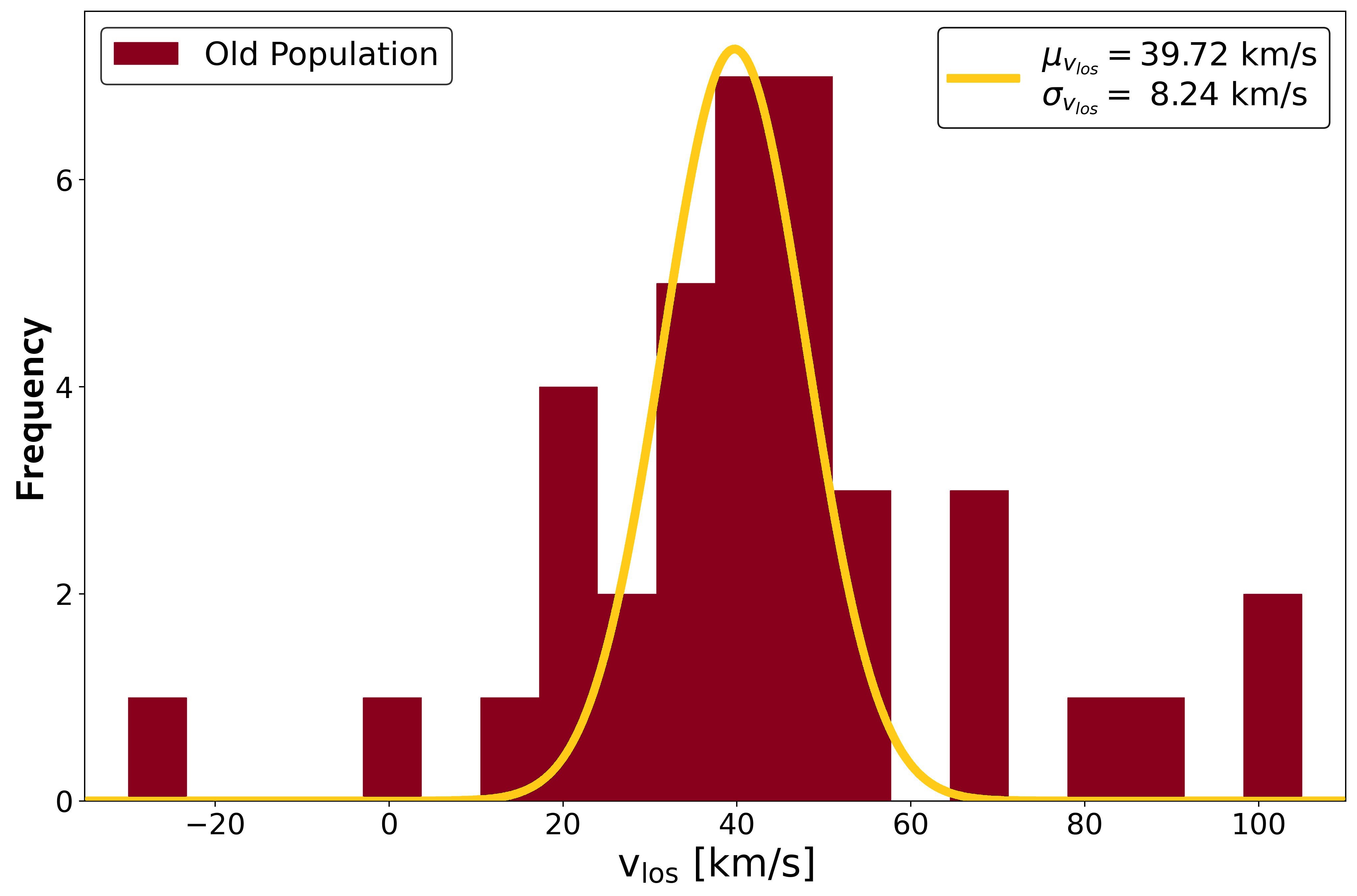}
  \caption{The distribution of the line-of-sight velocity of 55 old Leo T stars, estimated using \spexxy. Also plotted is the result of a MCMC fit for the mean and standard deviation of the distribution.}
  \label{hist_old}
\end{figure}

\subsection{Stellar Kinematics vs Neutral Gas Kinematics}
\begin{figure}[]
    \centering
  \includegraphics[scale=.44]{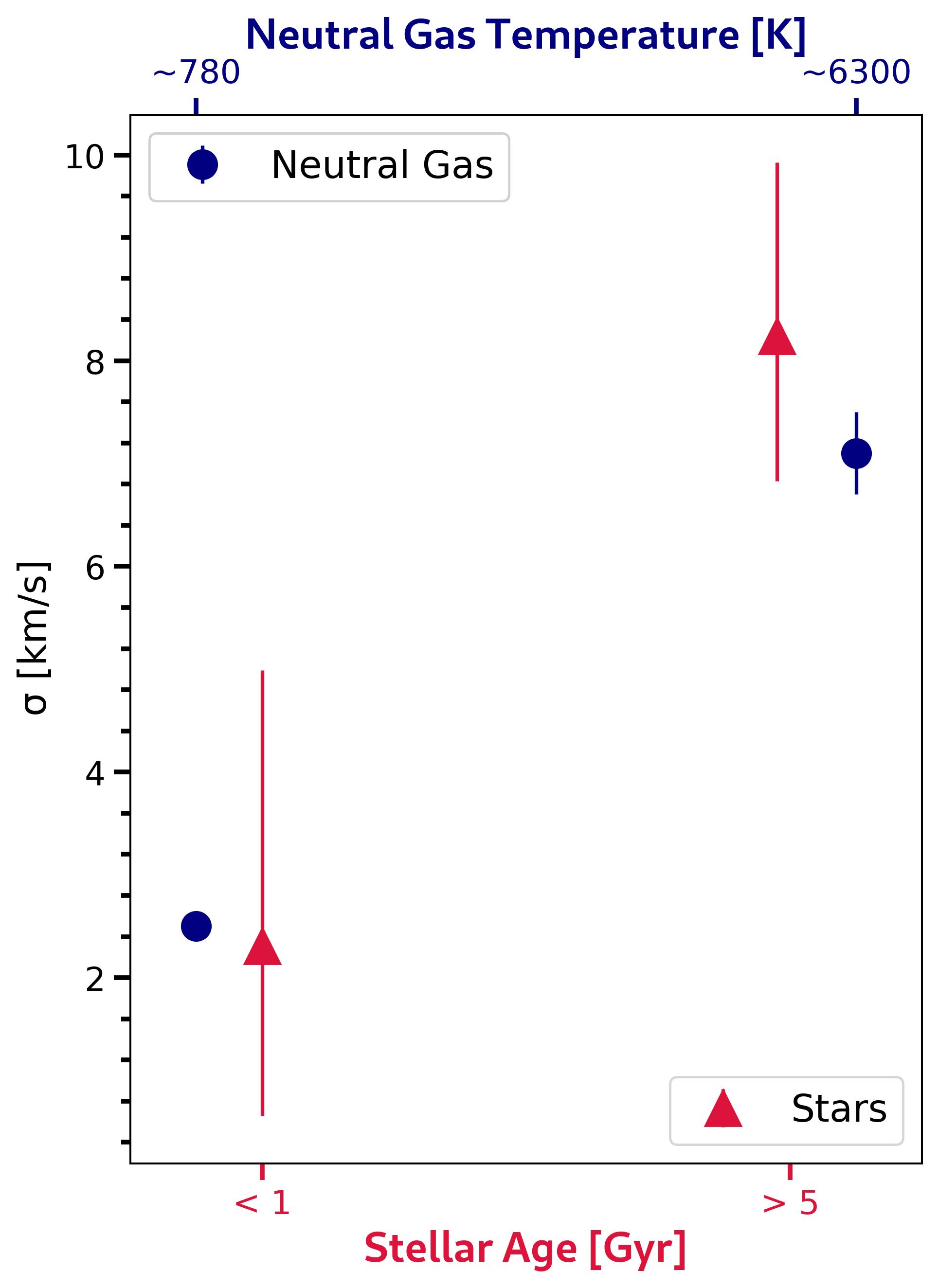}
  \caption{Comparison between stellar velocity dispersion of this work and neutral gas velocity dispersion from \cite{AdamsOosterloo2018}.}
  \label{gas_star}
\end{figure}

The histogram of the radial velocity estimates for 75 stars is shown in Figure~\ref{hist_vel}. To fit the distribution, we applied the same MCMC model as before to obtain the mean barycentric velocity and the velocity dispersion of the sample. The fit is also shown in Figure~\ref{hist_vel}. We find a mean velocity $v_{\mathrm{los}}=39.4^{+1.3}_{-1.3}\ \mathrm{km}\,\mathrm{s}^{-1}$, and an intrinsic velocity dispersion of $\sigma_{v} = 7.1^{+1.3}_{-1.1}\ \mathrm{km}\,\mathrm{s}^{-1}$, which is consistent with what was found by \cite{Simon2007}.
We repeated this analysis for each population. In this case, we used the same young population as before, but now the old population consists of 55 stars, including 17 stars from \cite{Simon2007}. These distributions and the respective fits are shown in Figures~\ref{hist_young} and \ref{hist_old} for the young and old population, respectively. It is worth noting that the best fit plotted does not include uncertainties.
For the younger population we obtain a mean velocity of $v_{\mathrm{los}} = 39.3^{+2.1}_{-2.1}\ \mathrm{km}\,\mathrm{s}^{-1}$, and a velocity dispersion of $\sigma_{v} = 2.3^{+2.7}_{-1.7}\ \mathrm{km}\mathrm{s}^{-1}$. For the older population we find a mean velocity of $v_{los} = 39.7^{+1.6}_{-1.6}\ \mathrm{km}\,\mathrm{s}^{-1}$, and a velocity dispersion of $\sigma_{v} = 8.2^{+1.7}_{-1.4}\ \mathrm{km}\,\mathrm{s}^{-1}$.
Notably, we find that both populations have different kinematics, with the younger population having a significantly smaller velocity dispersion than the older stars. 
This is comparable to what was found for the two components of the gas in Leo T, where the cold component was found to have a velocity dispersion smaller than that of the warm component. We compare the differences in kinematics between young and old stars with what was found for the \hi\ kinematics of warm and cold neutral gas in Figure~\ref{gas_star}. We find a good match when comparing the velocity dispersion
of the young population with the cold component of the \hi\ gas, and between the kinematics of old Leo T stars and the warm component of the \hi\ gas.
The natural inference from these results is that the most recent star formation in Leo T is linked to the CNM. 
The results presented here combined with the results from \cite{Weisz2012} of no star formation in Leo T in the last $\sim 25$ Myr are consistent with recent models that suggest that star formation in low mass galaxies should be bursty with short quiescent periods \citep{Collins2022}: due to stellar feedback, the star formation is momentarily quenched and metals are ejected from the environment and, after a short quiescent period, the gas is allowed to cool down and re-ignite star formation. 

\end{document}